\documentstyle[twoside,fleqn,espcrc2]{article}

\def\spose#1{\hbox to 0pt{#1\hss}}
\def\ltapprox{\mathrel{\spose{\lower 3pt\hbox{$\mathchar"218$}}
 \raise 2.0pt\hbox{$\mathchar"13C$}}}
\def\gtapprox{\mathrel{\spose{\lower 3pt\hbox{$\mathchar"218$}}
 \raise 2.0pt\hbox{$\mathchar"13E$}}}
\def\inapprox{\mathrel{\spose{\lower 3pt\hbox{$\mathchar"218$}}
 \raise 2.0pt\hbox{$\mathchar"232$}}}

\newcommand{\id}{\mbox{1$\!\!$I}}

\renewcommand{\Pr}{\hat{\mbox{I}\!\!\mbox{P}}}

\newcommand{\NPB}{Nucl. Ph. B}
\newcommand{\PLB}{Ph. Lett. B}
\newcommand{\PRD}{Ph. Rev. D}

\renewcommand{\>}{\rangle}

\title{Non-perturbative renormalization in kaon decays.}

\author{
A.~Donini\address{Dip. di Fisica, Universit\'a di Roma ``La Sapienza'' 
and I.N.F.N., Sezione di Roma I, \\ Piazzale Aldo Moro 2, I-00185 Roma, 
Italy.},
V.~Gim\'enez\address{Dep. de Fisica Teorica and IFIC,
Universitat de Val\`encia, E-46100 Burjassot (Valencia), Spain.}, 
G.~Martinelli$^{a}$,
G.C.~Rossi\address{Dip. di Fisica, Universit\'a di Roma ``Tor Vergata''
and I.N.F.N., Sezione di Roma II, \\ 
Via della Ricerca Scientifica 1, I-00133 Roma, Italy.},
M.~Talevi$^{a}$\thanks{Talk presented by M.~Talevi}, 
M.~Testa$^{a}$
and A.~Vladikas$^{c}$
}

\begin{document}

\begin{abstract}
We discuss the application of the MPSTV non-perturbative 
method \cite{NPM} to the operators relevant to kaon decays. 
This enables us to reappraise the long-standing question of the
$\Delta I=1/2$ rule, which involves power-divergent subtractions that cannot be 
evaluated in perturbation theory.  We also study the mixing with
dimension-six operators and discuss its implications to the chiral behaviour
of the $B_K$ parameter.
\end{abstract}

\maketitle

\section{Introduction}

A quantitative theoretical understanding of the $\Delta I=1/2$ rule in 
$K\to\pi\pi$ decays has proven to be a formidable task 
since the calculation of hadronic matrix 
elements in the low-energy non-perturbative regime is needed.
In the continuum, with an active charm quark and the GIM mechanism at work,
the operator basis given by
\begin{equation}
O^{\pm}_{LL}= \frac{1}{2}[(\bar s d)_L(\bar u u)_L
                      \pm (\bar s u)_L(\bar u d)_L]   -(u\to c).
\label{eq:OpmGIM}
\end{equation}
In the framework of lattice QCD with Wilson-like fermions,
the renormalization strategy is complicated by chiral symmetry breaking.
In fact, the Wilson term induces the mixing of $O^{\pm}_{LL}$ with 
lower-dimensional operators,
with power-divergent coefficients, which need to be subtracted 
non-perturbatively.  Non-perturbative (NP) renormalization has witnessed 
a great progress recently \cite{NPM,Rossi,ALPHA,JLQCD}.  
In this talk, we want to outline the 
strategy for the renormalization of the lattice operators $O^{\pm}_{LL}$
using the NP method (NPM) of \cite{NPM}.
In the following, we shall concentrate only on the octet component of 
$O^{\pm}_{LL}$, which we will denote with $O^{\pm}_0$ 
\cite{Maiani,DI_TH}\footnote{The lattice penguin operators, being 
proportional to $(m_c^2-m_u^2)a^2\ll 1$, will be neglected in the following.}. 
In sec.~\ref{sec:d=6} we also reconsider the mixing with dimension-six 
operators \cite{DS=2,NP_pm} 
and its implications on the chiral behaviour of $B_K$ \cite{B_K}, 
a problem also addressed in \cite{JLQCD} using the Ward Identities (WI). 
We stress that the NPM of \cite{NPM} and the WI 
are completely equivalent, provided the renormalization scale is large 
enough, since the renormalized operators with the good chiral properties are
unique.  Anyway, the NPM of \cite{NPM} must be 
used to calculate the overall costant, which the WI cannot determine.

\section{Renormalization strategy}

On the lattice, the general prescription to restore the chiral symmetry
broken by the Wilson term is to subtract all the operators of dimension
equal or less than $O^{\pm}_0$.  The operators which need to be subtracted
are dictated by the symmetries of the action:  charge conjugation (C),
parity (P) and $s\leftrightarrow d$ flavour switching symmmetry (S) \cite{BDHS}.

There are different ways of calculating the $K\to\pi\pi$ matrix elements,
which correspond to different renormalization structures \cite{Rossi,DI_TH}.
We considering here a general structure of the form
\begin{equation}
\widehat O^{\pm}=Z^{\pm}\left[O^{\pm}_0+\sum_{i=1}^4 Z^{\pm}_iO^{\pm}_i
+Z^{\pm}_5O_5+Z^{\pm}_3O_3\right]
\label{eq:hatOpm}
\end{equation}
where $O^{\pm}_0$ are the bare operators,
$O^{\pm}_i,\ i=1,\ldots,4$ are dimension-six operators of wrong 
chirality (cf.~sec.~\ref{sec:d=6}), $O_5$ is a dimension-five of the form 
$\bar s\Sigma_{\mu\nu}F_{\mu\nu}d$ ($\Sigma_{\mu\nu}=\sigma_{\mu\nu}$ or 
$\tilde\sigma_{\mu\nu}$) and $O_3$ is a dimension-three operator of the form
$\bar s\Gamma d$ ($\Gamma=\id$ or $\gamma_5$). 
According to the NPM, the mixing $Z$'s are determined by finding a set of 
projectors on the tree-level amputated Green functions (GF), with off-shell
quark and gluon external states,
the choice of which depends on the nature of the operators at hand.
For the $\Delta I=1/2$ operators we choose the following set of external 
states: $q\bar q$, $q\bar qg$, $q\bar qq\bar q$, with the momenta given 
below in eq.~(\ref{eq:6-Z-system}).  
For each choice of external states, i.e. for each different set of GF,
we need different type of projectors.  Let us denote with $\Pr_3$ 
the projector on the $q\bar q$ GF of the operator $O_3$, 
with $\Pr_5$ the projector on the $q\bar qg$ GF of the 
operator $O_5$, and with $\Pr_j,\ j=1,\ldots,4$ the set of mutually
orthogonal projectors on the operators $O_i,\ i=1,\ldots,4$ \cite{DI_TH,NP_pm}.
Applying the projectors to the corresponding NP GF
of the renormalized operators $\widehat O^{\pm}$,
with an appropriate choice of the external states, 
we require that the renormalized operators be proportional to the bare 
operators, $\widehat O^{\pm}(\mu)\propto O^{\pm}_0(a)$ (up to terms of 
${\cal O}(a)$), i.e.\ we impose the following renormalization conditions
(trace over colour and spin is understood in the projection operation):
\begin{equation}
\begin{array}{l}
\Pr_3\<q(p)|\widehat O^{\pm}|\bar q(p)\>=0  \\
\Pr_5\<q(p-k)g(k)|\widehat O^{\pm}|\bar q(p)\>=0  \\
\Pr^{\pm}_j\<q(p)\bar q(p)|\widehat O^{\pm}|q(p)\bar q(p)\>=0,\ j=1,\ldots,4  
\end{array}
\label{eq:6-Z-system}
\end{equation}
where $p$ and $k$ denote the momentum of the external quark and gluon legs.
The system of equations (\ref{eq:6-Z-system}) completely determines in a NP 
way the renormalization constants, as we have six conditions 
(non-homogeneous due to the matrix elements of $O^{\pm}_0$, 
cf.~eq.~(\ref{eq:hatOpm}))
in six unknown mixing constants, $Z^{\pm}_i,\ i=1,\ldots,4,Z^{\pm}_5,Z^{\pm}_3$.

Unfortunately, since solving eq.~(\ref{eq:6-Z-system}) involves
delicate cancellations between large contributions, it may very likely result
in a very noisy determination, even with large statistics.  
An equivalent strategy we can adopt is:\\
1.\ We introduce an intermediate subtraction for the dimension-five 
and -six operators
\begin{equation}
\begin{array}{l}
\bar O^{\pm}_i=O^{\pm}_i+C^{(\pm,i)}_3 O_3 ,\ i=0,\ldots,4, \\
\bar O_5=O_5+C^{(5)}_3 O_3,
\end{array}
\end{equation}
and determine the power-divergent mixing constants $C^{(\pm,i)}_3$ and 
$C^{(5)}_3$ by imposing 
\begin{equation}
\begin{array}{l}
\Pr_3\<q(p)|\bar O^{\pm}_i|\bar q(p)\> =0,\ i=0,\ldots,4, \\
\Pr_3\<q(p)|\bar O_5|\bar q(p)\> =0.
\end{array}
\end{equation}
2.\ The finite mixing constants $Z^{\pm}_i$ and $Z^{\pm}_5$,
which in principle can be calculated in perturbation theory (PT),
are then determined from the system 
\begin{equation}
\begin{array}{l}
\Pr_5\<q(p-k)g(k)|\sum_iZ^{\pm}_i\bar O^{\pm}_i
                 +Z^{\pm}_5\bar O_5|\bar q(p)\> \\
=\Pr_5\<q(p-k)g(k)|\bar O^{\pm}_0|\bar q(p)\> \\
\Pr_j\<q(p)\bar q(p)|\sum_iZ^{\pm}_i\bar O^{\pm}_i
                    +Z^{\pm}_5\bar O_5|q(p)\bar q(p)\> \\
=\Pr_j\<q(p)\bar q(p)|\bar O^{\pm}_0|q(p)\bar q(p)\>,\ j=1,\ldots,4.
\end{array}
\end{equation}

\section{Mixing with dimension-six operators}
\label{sec:d=6}

In ref.~\cite{DS=2} the NP mixing with equal dimensional operators of
the $\Delta S=2$ operator, which has the same Lorentz structure of $O^+_0$, 
has been studied, with the mixing limited to a
basis of three operators, denoted here with $O^+_i,\ i=1,2,3$, found
by an explicit 1-loop calculation \cite{PT_4f}.
A more careful analysis based on CPS\cite{NP_pm} has shown that beyond 1-loop, 
the operators $O^{\pm}_0$ are allowed to mix not only with 
$O^{\pm}_i,\ i=1,2,3$, present in 1-loop PT, but also with 
\begin{equation}
O^{\pm}_4\equiv\frac{(\pm N_c-1)}{16N_c}
        [ ( SS + PP -\frac{1}{3} TT ) \pm (2\leftrightarrow 4) ],
\label{eq:O_4}
\end{equation}
which is not present at the 1-loop level, where
$\Gamma\Gamma\equiv (\bar\psi_1\Gamma\psi_2)(\bar\psi_3\Gamma\psi_4)$
and $N_c=3$.

The operators $O^{\pm}_i,\ i=0,\ldots,4$ are Fierz eigenstates in Dirac-colour
space with eigenvalue $\pm 1$, respectively \cite{NP_pm}.
We stress that Fierz transformations are not a symmetry of the action,
but only of the operators, thus they cannot determine the operator mixing.  
Nevertheless, since Fierz transformations have been used in the perturbative
calculations \cite{PT_4f}, we have also used them to classify and reorganize 
the NP mixing.

\begin{figure}[t]   
    \begin{center}
       \setlength{\unitlength}{1mm}
       \begin{picture}(50,50)   
          \put(-15,-35){\includegraphics{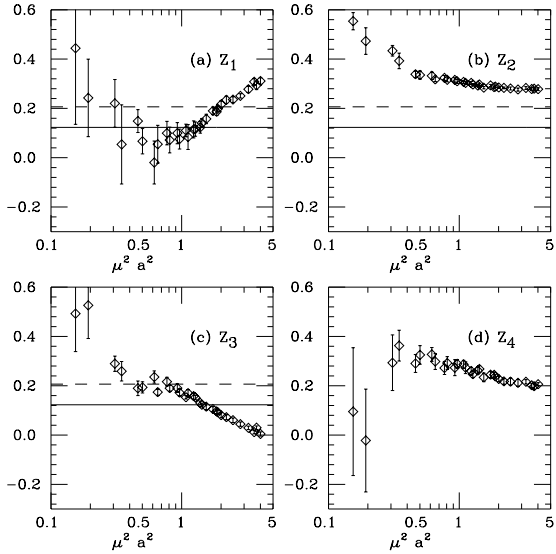}}   
       \end{picture}
    \end{center}
\caption{NP mixing $Z$'s as a function of $\mu^2a^2$.  
The solid (dashed) line is from SPT (BPT).}
\label{fig:zi}
\end{figure}
\begin{figure}[t]   
    \begin{center}
       \setlength{\unitlength}{1mm}
       \begin{picture}(50,50)   
          \put(-25,-40){\includegraphics{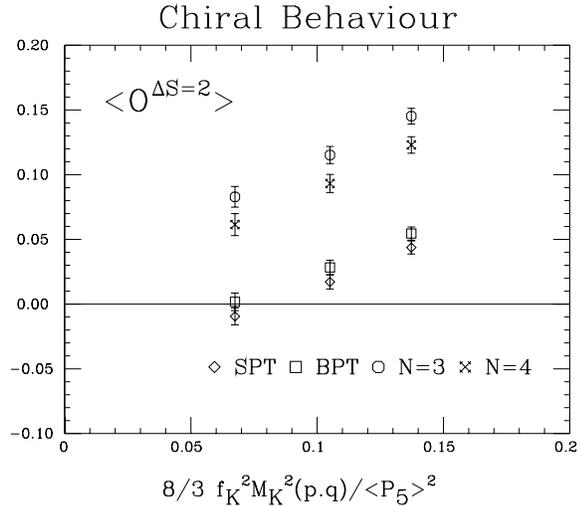}}   
       \end{picture}
    \end{center}
\caption{Chiral behaviour of $B_K$.  The NP $Z$'s are taken at $\mu^2a^2=0.96$.}
\label{fig:chiral}
\end{figure}

We have extended the method for the renormalization
of the $\Delta S=2$ operator presented in ref.~\cite{DS=2} including 
the fourth operator $O_4$, and have determined the mixing constants 
$Z_i,\ i=1,\ldots,4$, and the overall constant $Z^{\Delta S=2}$.  
We have performed the calculation using the same parameters of 
ref.~\cite{DS=2}, on a $16^3\times 32$ lattice, 
at $\beta=6.0$, with a hopping parameter $\kappa=0.1425$ 
for the SW-Clover quark propagator, in the lattice Landau gauge,
but with higher statistics, i.e.\  on an ensemble of 200 configurations.
We have calculated the $Z$'s at different renormalization scales $\mu^2a^2$ 
(cf.\ fig.\ \ref{fig:zi}).  
We note that $Z_2$ and $Z_4$ are very well defined and almost scale-independent
in a large ``window'' of $\mu^2a^2$, whereas $Z_1$ and $Z_3$
present a smaller window, i.e.\ a more pronounced scale-dependence.
Moreover, $Z_4$ which is absent in 1-loop PT, is not negligible.
We stress that the large fluctuations at small $\mu^2a^2$ do not 
spoil the validity of the NPM, since in that region the perturbative
matching to a continuum scheme is not reliable, as for any
NP lattice method \cite{NPM,DS=2}.

Using the NP $Z$'s of fig.~\ref{fig:zi}, the chiral behaviour 
of the $B_K$ parameter studied in refs.~\cite{DS=2,B_K} can be revisited.
Since the mixing with $O_4$ starts at ${\cal O}(g_0^4)$ we do not expect 
drastic changes in the chiral behaviour.  If the chiral behaviour were
sensibly different, we would not trust the matching to the continuum which
has an uneludable perturbative uncertainty. 
In fig.\ \ref{fig:chiral} the result using the NP $Z$'s at $\mu^2=.96$ is 
shown.  
We will concentrate here on the novelties introduced by the NP mixing with
the complete basis ($N=4$) with respect to the basis found in 1-loop PT 
($N=3$) \cite{DS=2,B_K}, refering to \cite{NP_pm} for a detailed analysis.
Using the usual parametrization
$\<O^{\Delta S=2}\>=\alpha+\beta m_K^2+\gamma(p\cdot q)+...$,
we have found that the use of the complete set of operators leaves the
values of $\beta$ and $\gamma$ almost unchanged compared to the $N=3$ case,
whereas the value of $\alpha$ is slightly lowered, 
passing from $\alpha^{N=3}=0.017(13)$ to $\alpha^{N=4}=-0.004(15)$.
We note that the two values of $\alpha$ are compatible with each other
and with zero.   
Within the current statistical limitations forced by the thinning
approximation \cite{B_K}, we can only state that 
the value of $B_K$, proportional to $\gamma$, is unaltered, 
while the correct chiral behaviour of the continuum,
signaled by the vanishing of $\alpha$, is recovered.  
These results are 
independent of the scale $\mu$ used, albeit not too small.
We argue that this stability is due to the extremely clean determination of 
$Z_2$ \cite{NP_pm}.

\end{document}